# The Umwelt Representation Hypothesis: Rethinking Universality

Victoria Bosch*[1], Rowan P. Sommers*[1], Adrien Doerig[2,3+], and Tim C. Kietzmann[1+]

[1] Institute of Cognitive Science, University of Osnabrück, Osnabrück, Germany.
[2] Department of Psychology and Education, Freie Universität Berlin, Berlin, Germany.
[3] Bernstein Center for Computational Neuroscience, Berlin, Germany
* shared first authorship
+ shared last authorship

**Correspondence:** vbosch@uni-osnabrueck.de, tim.kietzmann@uni-osnabrueck.de

**Keywords:** universality, computational modeling, artificial neural networks, representational alignment, neuroconnectionism

Recent studies reveal striking representational alignment between artificial neural networks (ANNs) and biological brains, leading to proposals that all sufficiently capable systems converge on universal representations of reality. Here, we argue that this claim of Universality is premature. We introduce the Umwelt Representation Hypothesis (URH), proposing that alignment arises not from convergence toward a single global optimum, but from overlap in ecological constraints under which systems develop. We review empirical evidence showing that representational differences between species, individuals, and ANNs are systematic and adaptive, which is difficult to reconcile with Universality. Finally, we reframe ANN model comparison as a method for mapping clusters of alignment in ecological constraint space rather than searching for a single optimal world model.

## 1. Are brains and ANNs converging to universal representations?

Animals navigate the complex world relying on neural representations that capture relevant aspects of their environment. To make progress in understanding these neural representations, artificial neural networks (ANNs) have become a prominent tool as they provide a language for expressing computational hypotheses about brain function [1–3]. A key approach to deriving biological insights from ANNs is to compare the degree to which models with different architectures, training objectives, datasets, and learning rules align with neural and behavioral data [1,3,4]. Using this approach, referred to as Neuroconnectionism, numerous studies have identified shared representational structure across brains and artificial neural network models, as well as factors influencing this alignment [5–11].

However, recent results suggest that contrasting different ANNs against neural data may start yielding diminishing returns: models with drastically different architectures and training regimes all match neural data with similar accuracy, especially when parametric flexibility is allowed for the model-brain mapping [6,12–14]. Furthermore, recent studies found a striking representational alignment between a wide variety of ANNs as well as the human brain, despite them having different architectures, network initialisations, learning objectives, and even different modalities such as language and vision [13,15–22].

These empirical findings have led to the theoretical proposal of representational Universality: different information-processing systems (ANNs and brains) all converge upon the same shared representational structures, because there exist "general-purpose dimensions that have little to do with the details of a network's architecture or task objective" [13], such that "all roads lead to Rome" (i.e., all systems learn the same representational structure) [23]. Universality is often seen as a consequence of ANNs and brains capturing a shared, statistical model of reality [13,16,24]. On this view, good biological and artificial systems converge toward faithful world representations, explaining their alignment (see Section 2; Fig. 1A).

Taken at face value, this theoretical explanation of the findings has important implications for neuroconnectionism. If diverse systems all converge on the same universal representations, we may no longer be able to gain neuroscientific insight by identifying which model ingredients lead to the most brain-like representations. Are we approaching the limits of using model comparison to study the brain? In this contribution, we first outline how recent work interprets representational alignment as convergence to a global optimality. We then provide an alternative interpretation introducing a framework grounded in the ecological concept of Umwelt. We then discuss how, despite the existence of shared representations, empirical evidence for systematic differences between perceptual systems contradicts strong claims of universality. Finally, we leverage this framework to reassess the role of model comparison in cognitive (neuro)science.

## 1. Universality presupposes global optimality

In order to explain why representations align, Universality appeals to normative theories of neural coding. These normative theories propose that cognitive systems converge on similar representations because they optimize the same normative goal and find similar solutions. A classic example of normative explanation pertains to why humans, mice, and various computational models all exhibit oriented edge detectors: these detectors are optimal for decomposing natural images under energy constraints [25,26]. What is the normative goal which Universality assumes to be driving representational alignment? Universality hypothesizes that certain representations are *globally optimal* for representing objective reality (i.e., the underlying data-generating process of 'the world'), and thus tend to emerge as a common solution. The best example of such a Universality position is the widely cited Platonic Representation Hypothesis [16], which draws from the philosophical tradition of metaphysical realism (see Box 1 for a brief overview of philosophical positions). It suggests that shared representations arise because "*models are recovering ever better representations of the actual world outside* [Plato's] *cave*", implying that there exists a single, globally optimal way of representing the world. Accordingly, all sufficiently capable systems, given enough capacity and data, should ultimately converge on the same representations of the world, despite differences in architecture, input modality, etc. Thus, we take representations to be *universally aligning* when *all* (well optimized) systems share this representation. In contrast, *partially aligned representations* are shared only by some systems (Fig. 1B).

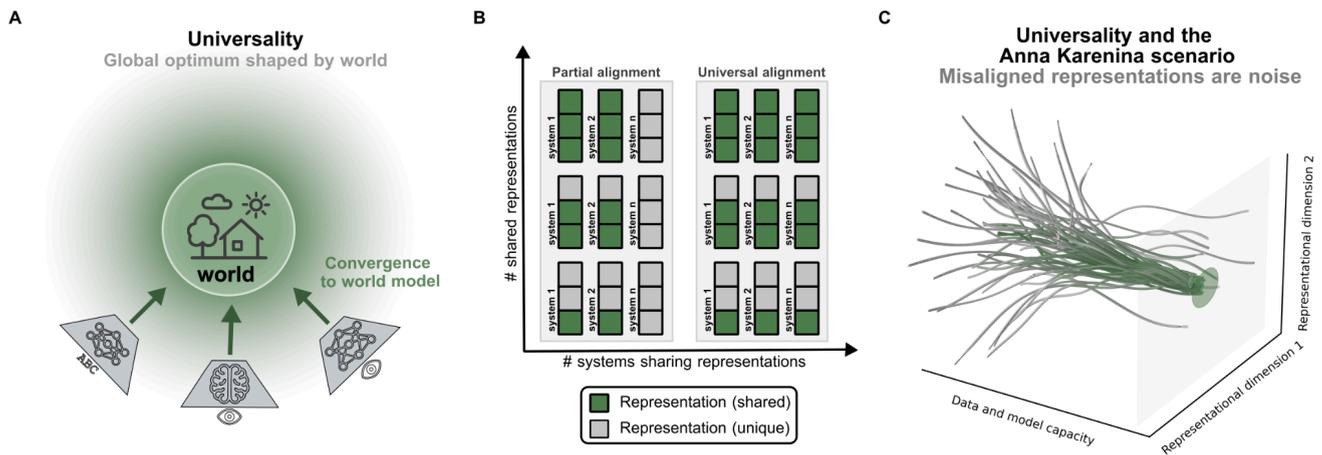

**Figure 1 | Universality. A.** Strong Universality claims such as the Platonic Representation Hypothesis postulate that different systems, regardless of input data, architectures, etc., converge to the same representation of the world (world model). **B.** Classification of representational alignment in terms of the number of systems that share representations, and how many representations are shared. We distinguish *universally aligned representations* (all well optimized systems share at least some representations) from *partially aligned representations* (some, but not all, systems share some, but not all, representations). **C.** Universality postulates that when we increase the data and model capacity of systems, their representations will converge to the same optimal point. The Anna Karenina scenario further suggests that representations close to the optimum are good (green), whereas representations further away from this optimum reflect idiosyncratic noise (gray).

Another interpretation in line with Universality is the 'Anna Karenina scenario' [22], after the famous opening line of Tolstoy's novel: "All happy families are alike; each unhappy family is unhappy in its own way." In the context of representational Universality, this principle can be extended from entire models to representational dimensions: dimensions that capture the true structure of the world should be shared across systems, whereas unshared dimensions are interpreted as idiosyncratic errors or non-optimal solutions [13,16,22,23]. Therefore, Universality seems to imply that all dissimilar representations reflect idiosyncratic noise (Fig. 1C).

> **Box 1. What World Should a World Model Represent?**
>
> The concept of a "world model", an internal representation that allows a system to predict and reason about its environment, has become central to machine learning. Yet it carries philosophical commitments involving a millennia-long debate on realism [84–86]. While resolving this debate is beyond the scope of this article, it is useful to ask: *what exactly is the target of a world model?*
>
> **The metaphysical realist view: world models as mirrors of nature.** From a metaphysical realist perspective [87], the goal of world models is to reach a correspondence to reality. The realist holds that the world possesses a unique observer-independent structure, and importantly, that an observer is capable of 'carving nature at its joints'. Accordingly, the Platonic Representation Hypothesis suggests that sufficiently powerful models trained on sufficiently rich data will converge on the same representation, because they are all approximating the same underlying reality. Metaphysical realism offers a parsimonious explanation for why independently trained systems develop similar internal structure.
>
> **The pragmatist alternative: world models as maps for action.** A contrasting view, drawing on pragmatist [88–90], phenomenologist [91,92] and post-structuralist traditions [93], holds that world models should be evaluated not by fidelity to a mind-independent reality but by their usefulness: a good representation captures the *differences that make a difference* for the system's goals [94,95]. A hunting tiger needs a model of prey location and action affordances, not a particle-level simulation. On this account, language and vision models develop similar representations not because both have discovered Platonic ideals, but because both face overlapping functional pressures: they must represent the same coarse-grained causal structure to perform well.
>
> **Pragmatism is more useful for cognitive neuroscientists.** Under realism, convergence should be global: deviations from a single target indicate incomplete learning (Anna Karenina scenario). Under pragmatism, convergence is *partial and task-relative*: models share structure where they face similar demands but may occupy a *multiplicity of equally valid optima* that differ because they serve different purposes. This maps onto the Umwelt concept: the world models of a bat and a bee share structure where their niches overlap but diverge where they do not, and neither is more "correct" in any absolute sense. Regardless of consensus amongst philosophers (or lack thereof), we propose that pragmatism is closer to the standard practice in cognitive (neuro)science, where we seek relationships between environmental factors, objectives, neural architectures, behavior and more, independently of whether they correspond to 'objective truth'.

## 2. The Umwelt representation hypothesis

An alternative account to why representational alignment occurs, yet one that does not necessarily treat dissimilarity as noise, can be derived from an ecological perspective grounded in the concept of Umwelt. The term "Umwelt" refers to the specific way particular organisms perceive the world, influenced by what they find important [27–29] (Fig. 2A). Thus, rather than accessing all available environmental information, systems operate within 'limited' Umwelts. For example, the bee's visible spectrum includes ultraviolet wavelengths but excludes red. This spectrum enables bees to see the iridescence of flowers carrying crucial information for their survival that is invisible to humans.

Accordingly, the Umwelt Representation Hypothesis (URH) states that the Umwelt of a system determines its representations: it inhabits its own particular slice of reality of what it can sense, understand and act upon, which is shaped by the system operating and developing under ecological constraints. Among these ecological constraints we include the organism's local environment, but also its sensory apparatus, effector capacities, its internal goals, (metabolic) resources, etc. [30–32] (Fig. 2B), many of which have been selected through evolution. Note that ANNs can also be characterised as having an Umwelt, determined by constraints such as architecture, training data, and learning objectives (Fig. 2C).

Under the URH, cognition is not a matter of reaching a global representational optimum, but rather of finding balanced compromises between the many constraints imposed that determine a system's Umwelt. For instance, greater perceptual resolution is not always beneficial, and could even be maladaptive due to constraints on energy, size or processing speed. Accordingly, the brain discards ecologically irrelevant information to achieve lossily compressed representations adaptive for its particular challenges [33–35]. For example, high-resolution vision may be harmful to detect a predator-shaped blur speeding towards you, as it is computationally expensive and likely too slow. In addition, it is not energy efficient: processing all visual input at foveal accuracy would require a brain an order of magnitude larger [36]. Despite these costs, high acuity can be critical for certain hunting styles in other ecological niches, explaining acuity differences between vision in predatory and prey animals. Similarly, ANNs trained to recognize dog breeds detect different image features than ANNs trained to detect tumors, but are not objectively better or worse world models; they have their own narrow niche. The URH thus challenges the idea that systems converge to 'one reality' with a unique global optimum towards which all systems strive.[1]

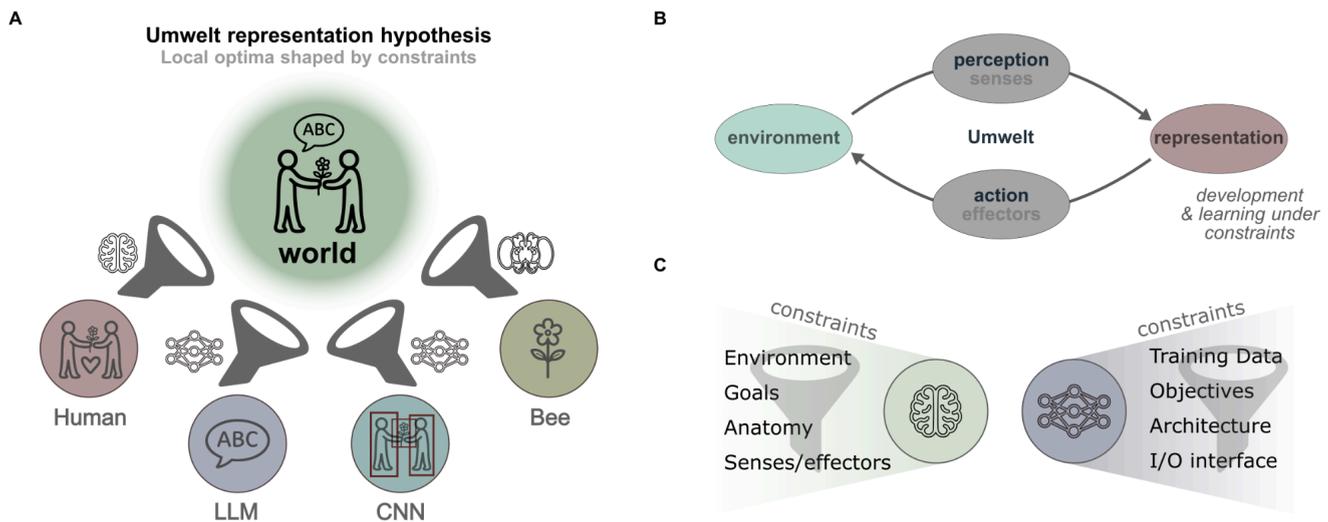

**Figure 2 | The Umwelt representation hypothesis. A.** The Umwelt representation hypothesis emphasizes how different systems (shaped under different constraints) view 'the world' through their own filter. These filters are themselves determined by the individual systems' ecological constraints (onto- and phylogenetic), and can result in both convergent and divergent representations, depending on constraint overlap. **B.** Umwelts are shaped by development and learning under constraints, through interaction of the system and the world mediated by sensors and effectors. **C.** Both human and ANN vision are shaped by analogous biological and artificial constraints. The content and effect of these constraints determine representational alignment between compared systems.

If, according to URH, the Umwelt of the individual determines the representations it learns, then why does representational alignment occur? According to the URH, similarities in input-output interfaces, objectives, and data statistics across systems give rise to representational alignment, while dissimilarities lead to non-aligned representations. This perspective naturally predicts *partial alignment*: some, but not all, systems share some, but not all, representations. For example, two human visual systems may align on coarse object distinctions, such as faces [37], but diverge in fine-grained identity information, while more distant systems, such as a fly, may share even fewer representations.

---

[1] There is a theoretical discussion to be had over the possibility that increasing the capacity and data of species will make differences between species disappear. We lean towards the view that even in such scenarios systematic differences would remain because of other ecological constraints, e.g. providing a bee with more brain capacity and data will not give it human representations. Likewise, there is a similar theoretical discussion over the possibility that massive scale in terms of data and capacity may enable a globally optimal foundation model which contains all Umwelts combined into one. But we predict that such a model will require some information selection mechanism to focus on the ecologically relevant regime, meaning this model—as a whole—will still not be a good model of individual systems; only subspaces inside this foundation model will. Regardless, we think neither of these theoretical discussions over possible scenarios are helpful in explaining the actual observable differences between species.

Importantly, not all ecological constraints need to fully overlap in order to develop similar solutions. Take convergent evolution, as an example: the photoreceptor system has evolved several times independently [38], illustrating that similar solutions can emerge when constraints overlap on a deeper structural level. Though this may seem to reintroduce a universal latent structure of the world, we emphasize that these deeper structures need not involve a globally optimal solution: they may still arise solely from the intersection between a few Umwelts. Nevertheless, the URH does not categorically deny the potential existence of universally shared representations either. Indeed, it is plausible that certain basic features are shared by all animals. For instance, even though echo-location, vision, tactile sensation and whisker sensation contain very different types of information, all of them can indicate the presence of solid, impassable objects. The relevant contrast between the URH and Universality claims is not about whether universal representations exist, but about their ubiquitousness and importance: Universality predicts that all systems converge to the same globally optimal universal representations; in contrast, the URH predicts that different systems optimize different ecological constraints. Universal representations, if any exist, are explained as overlap between these ecological constraints; not as a preordained tendency towards an optimal world model.

In summary, the URH maintains that different systems pick up different regularities in the world. Which regularities are represented depends on selective pressures from ecological constraints. This means the URH predicts partial alignment: shared representations between systems are explained as being well optimised for shared selective pressures, but non-shared representations are not necessarily "bad in their own way", they may instead be locally optimal for different ecological constraints.

### 3. Representational differences are systematic

To adjudicate between Universality and the URH, we turn to empirical evidence for representations that are *not* shared between systems. Universality predicts that dissimilar representations reflect idiosyncratic noise (all bad representations are bad in their own way). In contrast, the URH allows for systematic differences between systems as they can reflect representations that are well optimized for different selective pressures. That is, URH does not interpret all differences as due to noise, or a system's inability to model the world, but rather as systematic and meaningful.

*Systematic differences across species:* Representational differences between species are well established, revealing the extraordinarily varied solutions found by different organisms to survive. This applies within modalities (e.g. the visual neurons tuned to form a stellar compass in the Bogong moth, the sixteen types of photoreceptors in mantis shrimp compared to only three in humans, or the echolocation of dolphins and bats) and also to modalities that exist in some species, but are wholly absent in others (e.g. magnetoreception, the ability of migratory birds to detect Earth's magnetic field for navigation). These differences are not noise, but highly systematic, and reflect adaptation to different ecological niches. This is illustrated by the finding that the evolutionary hierarchy is reflected in functional connectivity patterns [39], implying that evolutionary distances between species predict representational similarity. Importantly, representational differences exist even between species that evolved in overlapping environments, showing that structured differences also depend on other constraints instead of merely reflecting different environmental statistics. Viewing this diversity through a lens of incomplete convergence towards a globally optimal world model misses what makes these unique evolutionary adaptations so useful for each species. Systematic representational differences between species are better seen as locally optimal solutions that contribute to species' survival under ecological constraints.

*Systematic differences between individuals:* Even within species, there are well known systematic differences between individuals [40–42] (note that there are also non-systematic inter-individual differences usually discarded as noise, which we leave aside here). For example, individuals from different cultures can be subject to different illusions or perceive different colors [43,44]. Within cultures, percepts can vary drastically too, as exemplified by the well-known case of #thedress [45]. These systematic differences in subjective percepts indicate differences in the underlying representational profiles. Structured individual differences have been observed at the neural level too. For example, humans have different representational spaces for familiar vs. unfamiliar faces and objects [46], and entirely new functionally defined brain regions may arise from experience with reading [47], Pokémon [48] or cars [49]. Similarly, cross-linguistic variation in syntactic structure, phonological systems, and semantic categorization necessitate language-specific representations that can capture the unique grammatical properties, sound patterns, and conceptual distinctions inherent to each language [50]. Some differences are innate, such as color blindness, tetrachromacy, or perfect pitch [51,52], while others are acquired through learning. In short, striking differences in both representation and behavior arise from diverse evolutionary, developmental, and environmental factors, suggesting that representational variation is geared towards local, adaptive optimality as opposed to imperfect global optimality.

*Systematic differences between humans and ANNs:* Despite impressive abilities to model neural activity, current ANNs still exhibit strong and systematic differences to human representations and resulting behavior [53–55]. A well-known example is that most classically trained ANN models of vision are texture-biased, while humans rely more on global shape cues to recognize objects visually [56,57]. Such differences are also reflected in representational alignment, as improved model performance does not necessarily lead to better brain alignment [58], and representational alignment in recent work rarely reaches the noise ceiling, implying that models fail to capture variance that is shared between people. Moreover, systematic individual differences also arise between ANNs when varying network depth, dropout and random initializations [53,59], and training on different tasks naturally yields different representations [8]. Together, these structured differences (both between humans and ANNs and across individual ANNs) are hard to explain under the Anna Karenina scenario of imperfect global optimality. Instead, they are more naturally explained as local optimality constrained by the networks' architectures, training data and objectives (see also Section 4).

In summary, while recent results have put the spotlight on the large amount of shared representations between perceptual systems, there is also strong evidence for *non*-shared representations within and between types of systems. Crucially, there is *structure* and *systematicity* in these representational differences, which are better interpreted as being *adaptive* specializations for the system that operate at ontogenetic (developmental) or phylogenetic (evolutionary) scales. The perspective of global optimality is not well-suited to account for the coexistence of alignment and systematic differences, because it misclassifies differences as idiosyncratic errors or incomplete convergence, rather than recognizing them as outcomes of adaptation and learning under local constraints. Even if there were indeed a global optimum to which all systems try to converge, but fail to due to limited capacity, the Universality perspective would still be unable to explain functionally relevant adaptations. Therefore, current evidence is better aligned with the URH than with Universality.

## 4. Model comparison as a method for probing constraint space

We started this piece by stating that model contrasting may be misguided under the Universality hypothesis, as all systems are predicted to, in the limit, converge on the same solutions. Does taking the Umwelt perspective salvage this approach, and, importantly, how does URH explain the striking degree of representational alignment observed in recent studies?

### 4.1 Alignment clusters in ecological constraint space

One way to explain the alignment is to highlight that despite seemingly large differences, the models studied all have ecological constraints that overlap on a deeper structural level: they all inhabit a human-centric Umwelt. Visual models are exposed to images of typical human environments, taken by humans. We train them to learn object categories that humans have named, or apply unsupervised learning tailored to yield representational spaces suitable for human-relevant tasks. Moreover, common model architectures have human-inspired inductive biases, such as hierarchical organisation, distributed coding, or retinotopic organisation. Thus, rather than convergence to universal representations, recent alignment studies have convincingly shown that the above ANNs cluster together with human brains. But since they have sampled mostly from human-centric ecological constraints, it remains to be seen whether the similarities are a consequence of narrow sampling in the space of possible model constraints.

This can be directly tested by determining the boundaries of the alignment cluster: what ecological constraints determine whether systems align or not? Does representational alignment optimally cluster around biological features alone or do we require intra-human factors such as cultural, political, cross-linguistic and personal developmental differences as well? To find out, we need to sample more broadly and measure representational alignment of models trained on datasets, architectures, etc. that diverge more strongly. If current ANNs indeed only align with humans and not with other organisms, then these representations are not *universally* aligned, but instead show that neuroscience and machine learning have discovered one particular cluster of partially aligning representations adaptive for a human-like Umwelt.

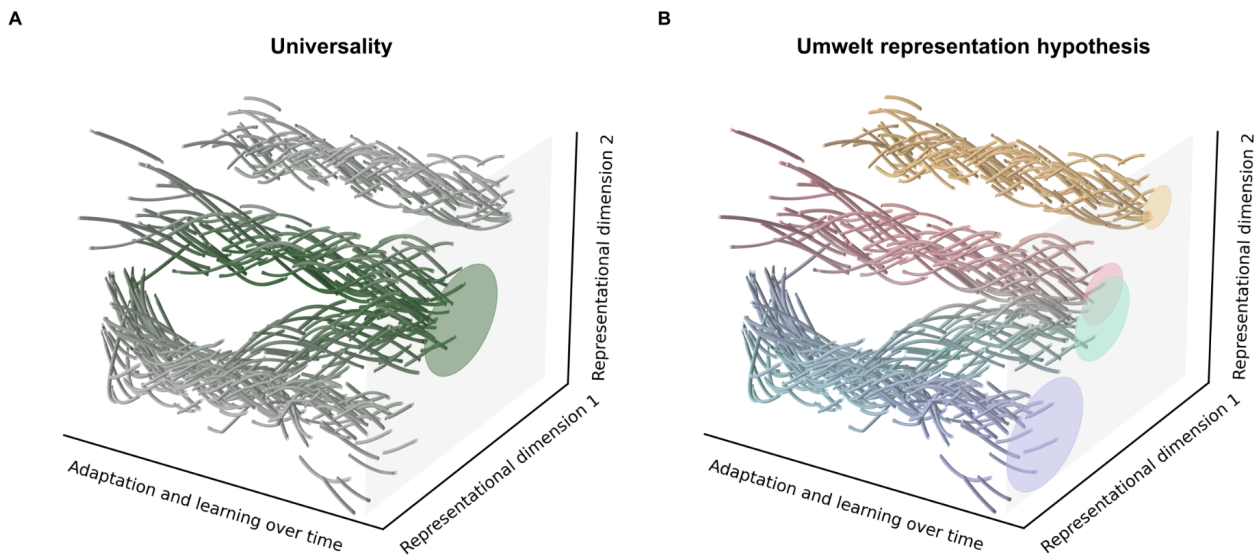

**Figure 3 | Universality vs. the Umwelt representation hypothesis as clustering hypotheses in representational space.** Illustrated here are two interpretations of the same simulated data depicting systematic similarities as well as differences between representations. Simulated trajectories show how different systems move through representational space with adaptation and learning over time (evolution/cultural change/development). A question for neuroscience would be to optimally identify clusters of systems reaching similar representations under shared ecological constraints. The shaded circles on the plane spanning the representational dimensions indicate potential clusters of systems in representational space. **A.** When interpreting this data through the lens of Universality, there is but one single cluster: those systems that arrive at a veridical world model (green trajectories depict "universal"/optimal systems; gray trajectories outside of the single cluster are suboptimal and reflect idiosyncratic noise). **B.** When interpreting the same simulated data via the Umwelt perspective multiple different clusters, each with their own locally optimal representations shaped by ecological constraints. Over time, clusters may converge and diverge (e.g. the teal cluster initially diverged from the purple one, and later converged to the red cluster).

Contrary to Universality, under the URH, we should not expect all systems to converge to the same global optimum. Instead, similar ecological constraints form local clusters of similar representations (see Fig. 3). This shifts the aim of neuroscience away from finding a single best optimal model that explains all cognitive systems, to finding clusters of systems that reach similar representations under similar ecological constraints. The resulting approach is in line with comparative systems neuroscience, allowing for identification of hierarchies and superclasses of similar Umwelts. Which types of ANN models cluster together with the representations of humans? Which ones cluster together with mice? How much do the human and mouse clusters overlap?

### 4.2 Exploring the space of ecological constraints

The above considerations raise the possibility that past findings of universality are limited to cases in which systems share a similar (human-centric) Umwelt. To expand this narrow view and find out which constraints matter for alignment, why and to what extent, we need to further explore the space of ecological constraints in search of alignment clusters, as well as work towards improving the sensitivity of alignment metrics (see Box 2).

*Model architectures:* The architectures explored in Universality studies are currently dominated by CNNs and transformers, which share strong inductive biases. Alternative architectures such as recurrent [60,61], topographic [62–64], probabilistic [65], predictive coding [66], or multimodal networks [67] remain relatively underexplored and may reveal new alignment clusters.

*Model training datasets:* Most studies rely on highly similar data distributions. Thus far, mainly human-derived vision and language datasets have been explored. Moreover, within explored modalities, dataset variety is still low. For example, most studies on Universality in vision have relied on datasets such as ImageNet and COCO [68,69] with relatively low resolution RGB static images representing human environments, and reflecting human visual biases, e.g. photographer bias. Greater dataset variety might reveal differences that are unobservable using the currently explored datasets. This possibility is supported by recent work showing that ANNs achieving high neural prediction scores on natural scenes fail on out-of-distribution synthetic images [70], and that targeted stimulus selection is a potent tool for differentiating between systems [22,55,71]. Moreover, investigating other species and underexplored modalities (e.g. olfaction, echolocation) perhaps has great potential to reveal model properties that distinguish alignment clusters in more distant and unexplored parts of ecological constraint space.

*Model objectives:* Models are mostly trained on object classification, or a small set of unsupervised objectives, hence important aspects of perception are potentially missed [72,73]. It is also found that models trained with different objectives best match different brain regions, indicating that not all objectives lead to the same representations [74]. Hence, we need to further explore possible learning objectives and ask how they shape representations.

*Further model constraints:* Perceptual systems operate under specific implementational and functional constraints, such as energetic efficiency, wiring restrictions, how rapidly capacities are acquired during development, and more. These constraints shape the brain's Umwelt and thereby which representations are learnt. However, they are largely absent from most models. For example, CNNs have a specific wiring pattern where weights are "copy-pasted" across layers. The brain cannot implement this, imposing a strong under-studied constraint. Creating non-convolutional models [63] is an exciting research direction, albeit bringing new challenges to the table, such as dealing with higher parameter counts. Finally, embedding artificial systems inside actual bodies provides additional constraints and affordances that may enable different representations to emerge [75–77]. Understanding and solving these challenges will help determine the constraints that the brain operates under and potentially exhibit new alignment clusters.

*Cognitive tasks:* Aside from the space of models tested in alignment studies, the space of cognitive tasks in cognitive (neuro)science remains sparsely sampled too. Recent state of the art large-scale neural datasets are primarily focussed on stimulus-locked static viewing, listening or reading paradigms. For instance, the Natural Scenes Dataset [78], a widely used open fMRI dataset with beautifully high signal-to-noise, consists of stimuli that originate from relatively few semantic clusters while human observers retain central fixation [79]. This is different from active vision, during which the activity profiles along the visual system may differ drastically [80–82]. Key perceptual and cognitive processes, such as learning dynamics, multi-object representations, memory, dynamic vision, and multimodal integration could likely shape alignment clusters too.

In sum, under the URH, model comparison is not undermined by recent findings of representational alignment between a subset of ANNs and brains on passive viewing of natural scenes. On the contrary, it remains an essential method for identifying the ecological constraints that shape when and why systems align, and when they diverge. Model comparison under the URH shifts from finding a single best world model, to finding clusters of models that share representations shaped by similar ecological constraints.

> **Box 2. Improving alignment metrics by increasing sensitivity to (mis)alignment.**
>
> A crucial aspect in discussions of representational alignment between perceptual systems is how sensitive our alignment metrics are: if our alignment metrics are not sufficiently sensitive, systems may seem aligned despite differing in important ways. Parametric flexibility of the metric is an important component in this respect. For example, commonly used encoding models fit a large number of parameters, affording substantial flexibility in how representations of one system are mapped onto another. While this can reveal relevant similarities between systems, it can also obscure important differences [96–98], limiting our ability to determine which model ingredients matter for representational alignment. As another example, a common approach across the field to quantify alignment is reporting average scores, such as variance explained, across large parts of the brain, or across stimuli. This too may mask differences between models, as they can account for different variance components (or predict responses to different stimuli well). Attempts of sharpening evaluation by combining metrics have been reported [54,99,100]. Better understanding the properties of existing metrics and developing new ones will sharpen our observations of what is shared or not between perceptual systems, which is essential for interpreting the results of model comparisons.

## 5. Concluding remarks

Recent findings of representational alignment between ANNs and brains have been explained by the proposal that they share the goal of accurately modelling the world, and therefore reach "optimal" universal representations. Here, we dissected the ideas behind Universality and evaluated empirical evidence supporting it. We argued that current evidence does not warrant such strong claims of Universality because there are structured, systematic, and adaptive differences between species, between individuals and between current ANNs and human brains. This challenges the Universalist explanation that attributes alignment to convergence on a single world model. Instead, we proposed an ecological reframing grounded in the concept of the Umwelt. According to this view, cognitive systems do not aim to achieve faithful world representations, but instead develop "locally optimal" representations to accommodate their needs to act in the world within the bodily and environmental constraints they are under. This perspective can inform neural network modelling and a more biology-inspired cognitive computational neuroscience [83]. Methodologically, this reframing re-emphasizes the importance of model comparison as a method to understand cognition by identifying clusters of representational alignment between models and brains, and explaining these clusters in terms of ecological constraints.

**Acknowledgements**

This work was funded by European Research Council's (ERC) Starting grant #101039524 "TIME" (VB, TCK). We want to thank Holger Lyre for helpful discussions at the project onset.

**Declaration of generative AI and AI-assisted technologies in the writing process.**

During the preparation of this work the authors used Claude Opus and ChatGPT 5.2 for aiding improvement of select sentences, as one source of information for validation of select references, and as coding assistant for generating Figure 1C and Figure 3. After using these tools, the authors reviewed and edited the content as needed and take full responsibility for the content of the published article.

# Bibliography

1.  Doerig, A. *et al.* (2023) The neuroconnectionist research programme. *Nat. Rev. Neurosci.* 24, 431–450
2.  Kietzmann, T.C. *et al.* (2018) Deep Neural Networks in Computational Neuroscience. *bioRxiv* DOI: 10.1101/133504
3.  Richards, B.A. *et al.* (2019) A deep learning framework for neuroscience. *Nat. Neurosci.* 22, 1761–1770
4.  Golan, T. *et al.* (2023) Deep neural networks are not a single hypothesis but a language for expressing computational hypotheses. *Behav. Brain Sci.* 46, e392
5.  Cichy, R.M. *et al.* (2016) Comparison of deep neural networks to spatio-temporal cortical dynamics of human visual object recognition reveals hierarchical correspondence. *Sci. Rep.* 6, 27755
6.  Conwell, C. *et al.* (2024) A large-scale examination of inductive biases shaping high-level visual representation in brains and machines. *Nat. Commun.* 15, 9383
7.  Dobs, K. *et al.* (2022) Brain-like functional specialization emerges spontaneously in deep neural networks. *Sci. Adv.* 8, eabl8913
8.  Dwivedi, K. *et al.* (2021) Unveiling functions of the visual cortex using task-specific deep neural networks. *PLOS Comput. Biol.* 17, e1009267
9.  Güçlü, U. and Gerven, M.A.J. van (2015) Deep Neural Networks Reveal a Gradient in the Complexity of Neural Representations across the Ventral Stream. *J. Neurosci.* 35, 10005–10014
10. Khaligh-Razavi, S.-M. and Kriegeskorte, N. (2014) Deep Supervised, but Not Unsupervised, Models May Explain IT Cortical Representation. *PLoS Comput. Biol.* 10, e1003915
11. Storrs, K.R. *et al.* (2021) Diverse Deep Neural Networks All Predict Human Inferior Temporal Cortex Well, After Training and Fitting. *J. Cogn. Neurosci.* 33, 2044–2064
12. Schrimpf, M. *et al.* (2018) Brain-Score: Which Artificial Neural Network for Object Recognition is most Brain-Like? *Neuroscience*.
13. Chen, Z. and Bonner, M.F. (2025) Universal dimensions of visual representation. *Sci. Adv.* 11, eadw7697
14. Storrs, K.R. *et al.* (2020) Noise ceiling on the crossvalidated performance of reweighted models of representational dissimilarity: Addendum to Khaligh-Razavi & Kriegeskorte (2014). *bioRxiv*.
15. Bonner, M.F. and Epstein, R.A. (2021) Object representations in the human brain reflect the co-occurrence statistics of vision and language. *Nat. Commun.* 12, 4081
16. Huh, M. *et al.* (2024) The Platonic Representation Hypothesis. *arXiv*.
17. Li, Q. *et al.* (2024) Representations and generalization in artificial and brain neural networks. *Proc. Natl. Acad. Sci.* 121, e2311805121
18. Sorscher, B. *et al.* (2022) Neural representational geometry underlies few-shot concept learning. *Proc. Natl. Acad. Sci.* 119, e2200800119
19. Liu, H. *et al.* (2023) Visual Instruction Tuning. *arXiv*.
20. Lu, K. *et al.* (2021) Pretrained Transformers as Universal Computation Engines. *arXiv*.
21. Merullo, J. *et al.* (2023) Linearly Mapping from Image to Text Space. *arXiv*.
22. Hosseini, E. *et al.* (2024) Universality of representation in biological and artificial neural networks. *bioRxiv*.
23. Bansal, Y. *et al.* (2021) Revisiting Model Stitching to Compare Neural Representations. *arXiv*.
24. Khosla, M. *et al.* (2024) Privileged representational axes in biological and artificial neural networks. *bioRxiv*.
25. Olshausen, B.A. and Field, D.J. (1996) Emergence of simple-cell receptive field properties by learning a sparse code for natural images. *Nature* 381, 607–609
26. Rao, R.P.N. and Ballard, D.H. (1999) Predictive coding in the visual cortex: a functional interpretation of some extra-classical receptive-field effects. *Nat. Neurosci.* 2, 79–87
27. Von Uexküll, J. (1992) A stroll through the worlds of animals and men: A picture book of invisible worlds. *Semiotica* 89
28. Uexküll, J. von (1909) *Umwelt und Innenwelt der Tiere*, Julius Springer
29. Thompson, E. (2007) *Mind in life: Biology, phenomenology, and the sciences of mind*, Belknap Press/Harvard University Press
30. Shapiro, L. and Spaulding, S. (2025) Embodied Cognition. In *The Stanford Encyclopedia of*

*Philosophy* (Summer 2025.) (Zalta, E. N. and Nodelman, U., eds), Metaphysics Research Lab, Stanford University
31. Matheson, H.E. and Barsalou, L.W. (2018) Embodiment and Grounding in Cognitive Neuroscience. In *Stevens' Handbook of Experimental Psychology and Cognitive Neuroscience* ((1st edn) ) (Wixted, J. T., ed), pp. 1–27, Wiley
32. Varela, F.J. *et al.* (2017) *The Embodied Mind, revised edition: Cognitive Science and Human Experience*, MIT Press
33. Balas, B. *et al.* (2009) A summary-statistic representation in peripheral vision explains visual crowding. *J. Vis.* 9, 13.1-1318
34. Bar, M. *et al.* (2006) Top-down facilitation of visual recognition. *Proc. Natl. Acad. Sci.* 103, 449–454
35. Greene, M.R. and Oliva, A. (2009) Recognition of natural scenes from global properties: Seeing the forest without representing the trees. *Cognit. Psychol.* 58, 137–176
36. Duncan, R.O. and Boynton, G.M. (2003) Cortical Magnification within Human Primary Visual Cortex Correlates with Acuity Thresholds. *Neuron* 38, 659–671
37. Konkle, T. and Caramazza, A. (2013) Tripartite Organization of the Ventral Stream by Animacy and Object Size. *J. Neurosci.* 33, 10235–10242
38. Nilsson, D.-E. (2013) Eye evolution and its functional basis. *Vis. Neurosci.* 30, 5–20
39. Xu, T. *et al.* (2020) Cross-species functional alignment reveals evolutionary hierarchy within the connectome. *NeuroImage* 223, 117346
40. Borovska, P. and de Haas, B. (2024) Individual gaze shapes diverging neural representations. *Proc. Natl. Acad. Sci.* 121, e2405602121
41. Feilong, M. *et al.* (2018) Reliable individual differences in fine-grained cortical functional architecture. *NeuroImage* 183, 375–386
42. Han, C. and Bonner, M.F. (2025) High-dimensional structure underlying individual differences in naturalistic visual experience. *arXiv*.
43. Segall, M.H. *et al.* (1963) Cultural differences in the perception of geometric illusions. *Science* 139, 769–771
44. Amir, D. and Firestone, C. (2025) Is visual perception WEIRD? The Müller-Lyer illusion and the Cultural Byproduct Hypothesis
45. Drissi Daoudi, L. *et al.* (2017) The role of one-shot learning in #TheDress. *J. Vis.* 17, 15
46. Charest, I. and Kriegeskorte, N. (2015) The brain of the beholder: honouring individual representational idiosyncrasies. *Lang. Cogn. Neurosci.* 30, 367–379
47. Dehaene-Lambertz, G. *et al.* (2018) The emergence of the visual word form: Longitudinal evolution of category-specific ventral visual areas during reading acquisition. *PLoS Biol.* 16, e2004103
48. Gomez, J. *et al.* (2019) Extensive childhood experience with Pokémon suggests eccentricity drives organization of visual cortex. *Nat. Hum. Behav.* 3, 611–624
49. Gilaie-Dotan, S. *et al.* (2012) Neuroanatomical correlates of visual car expertise. *NeuroImage* 62, 147–153
50. Evans, N. and Levinson, S.C. (2009) The myth of language universals: Language diversity and its importance for cognitive science. *Behav. Brain Sci.* 32, 429–448
51. Jordan, G. and Mollon, J. (2019) Tetrachromacy: the mysterious case of extra-ordinary color vision. *Curr. Opin. Behav. Sci.* 30, 130–134
52. Takeuchi, A.H. and Hulse, S.H. (1993) Absolute pitch. *Psychol. Bull.* 113, 345–361
53. Feather, J. *et al.* (2023) Model metamers reveal divergent invariances between biological and artificial neural networks. *Nat. Neurosci.* 26, 2017–2034
54. Feather, J. *et al.* (2025) Brain-Model Evaluations Need the NeuroAI Turing Test. *arXiv*.
55. Golan, T. *et al.* (2020) Controversial stimuli: Pitting neural networks against each other as models of human cognition. *Proc. Natl. Acad. Sci.* 117, 29330–29337
56. Geirhos, R. *et al.* (2021) Partial success in closing the gap between human and machine vision. *arXiv*.
57. Lu, Z. *et al.* (2025) Adopting a human developmental visual diet yields robust, shape-based AI vision. *arXiv*.
58. Linsley, D. *et al.* (2025) Better artificial intelligence does not mean better models of biology. *Trends Cogn. Sci.* DOI: 10.1016/j.tics.2025.11.016
59. Mehrer, J. *et al.* (2020) Individual differences among deep neural network models. *Nat. Commun.*

11, 5725
60. Kar, K. *et al.* (2019) Evidence that recurrent circuits are critical to the ventral stream's execution of core object recognition behavior. *Nat. Neurosci.* 22, 974–983
61. Kietzmann, T.C. *et al.* (2019) Recurrence is required to capture the representational dynamics of the human visual system. *Proc. Natl. Acad. Sci.* 116, 21854–21863
62. Blauch, N.M. *et al.* (2022) A connectivity-constrained computational account of topographic organization in primate high-level visual cortex. *Proc. Natl. Acad. Sci.* 119, e2112566119
63. Lu, Z. *et al.* (2025) End-to-end topographic networks as models of cortical map formation and human visual behaviour. *Nat. Hum. Behav.* DOI: 10.1038/s41562-025-02220-7
64. Margalit, E. *et al.* (2024) A unifying framework for functional organization in early and higher ventral visual cortex. *Neuron* DOI: 10.1016/j.neuron.2024.04.018
65. Kingma, D.P. and Welling, M. (2022) Auto-Encoding Variational Bayes. *arXiv*.
66. Millidge, B. *et al.* (2022) Predictive Coding: a Theoretical and Experimental Review. *arXiv*.
67. Hu, Y. and Mohsenzadeh, Y. (2025) Neural processing of naturalistic audiovisual events in space and time. *Commun. Biol.* 8, 110
68. Lin, T.-Y. *et al.* (2015) Microsoft COCO: Common Objects in Context. *arXiv*.
69. Russakovsky, O. *et al.* (2015) ImageNet Large Scale Visual Recognition Challenge. *arXiv*.
70. Gifford, A.T. *et al.* (2026) A 7T fMRI dataset of synthetic images for out-of-distribution modeling of vision. *Nat. Commun.* 17, 1589
71. Golan, T. *et al.* (2023) Testing the limits of natural language models for predicting human language judgements. *Nat. Mach. Intell.* 5, 952–964
72. Garcia, K. *et al.* (2024) Modeling dynamic social vision highlights gaps between deep learning and humans. *OSF*.
73. Doerig, A. *et al.* (2025) High-level visual representations in the human brain are aligned with large language models. *Nat. Mach. Intell.* 7, 1220–1234
74. Bartnik, C.G. *et al.* (2025) Representation of locomotive action affordances in human behavior, brains, and deep neural networks. *Proc. Natl. Acad. Sci.* 122, e2414005122
75. Thelen, E. and Smith, L.B. (1994) *A Dynamic Systems Approach to the Development of Cognition and Action*, The MIT Press
76. Gibson, J.J. (1979) *The Ecological Approach to Visual Perception: Classic Edition*, Houghton Mifflin
77. O'Regan, J.K. and Noë, A. (2001) A sensorimotor account of vision and visual consciousness. *Behav. Brain Sci.* 24, 939–973; discussion 973-1031
78. Allen, E.J. *et al.* (2022) A massive 7T fMRI dataset to bridge cognitive neuroscience and artificial intelligence. *Nat. Neurosci.* 25, 116–126
79. Shirakawa, K. *et al.* (2025) Spurious reconstruction from brain activity. *Neural Netw.* 190, 107515
80. Hayhoe, M. and Ballard, D. (2005) Eye movements in natural behavior. *Trends Cogn. Sci.* 9, 188–194
81. Amme, C. *et al.* (2024) Saccade onset, not fixation onset, best explains early responses across the human visual cortex during naturalistic vision. *bioRxiv*.
82. König, P. *et al.* (2016) Eye movements as a window to cognitive processes. *J. Eye Mov. Res.* 9
83. Carvalho, W. and Lampinen, A. (2025) Naturalistic Computational Cognitive Science: Towards generalizable models and theories that capture the full range of natural behavior. *arXiv*.
84. Chakravartty, A. *et al.* (2025) Scientific Realism. In *The Stanford Encyclopedia of Philosophy* (Winter 2025.) (Zalta, E. N. and Nodelman, U., eds), Metaphysics Research Lab, Stanford University
85. Miller, A. (2024) Realism. In *The Stanford Encyclopedia of Philosophy* (Summer 2024.) (Zalta, E. N. and Nodelman, U., eds), Metaphysics Research Lab, Stanford University
86. Khlentzos, D. (2025) Challenges to Metaphysical Realism. In *The Stanford Encyclopedia of Philosophy* (Fall 2025.) (Zalta, E. N. and Nodelman, U., eds), Metaphysics Research Lab, Stanford University
87. Sider, T. (2011) *Writing the Book of the World*, Oxford University Press
88. James, W. (2019) Pragmatism - A New Name for Some Old Ways of Thinking by William James : With a Critical Introduction by Eric C. Sheffield. at <https://www.torrossa.com/en/resources/an/5671443>
89. Dewey, J. (1930) The Quest for Certainty: A Study of the Relation of Knowledge and Action. *J. Philos.* 27, 14–25

90. Rorty, R. (1979) *Philosophy and the mirror of nature*, Princeton University Press
91. Merleau-Ponty, M. (2006) *Phenomenology of perception: an introduction*, (Repr.), Routledge
92. Heidegger, M. (2010) *Being and Time*, SUNY Press
93. Deleuze, G. and Guattari, F. (1987) *A Thousand Plateaus*, University of Minnesota Press
94. Bateson, G. (2000) *Steps to an Ecology of Mind: Collected Essays in Anthropology, Psychiatry, Evolution, and Epistemology*, University of Chicago Press
95. Dennett, D.C. (2017) *From Bacteria to Bach and Back: The Evolution of Minds*, W. W. Norton & Company
96. Avitan, I. and Golan, T. (2025) Rethinking Representational Alignment: Linear Probing Fails to Identify the Ground-Truth Model. *Cogn. Comput. Neurosci. CCN*
97. Thobani, I. *et al.* (2025) Model-brain comparison using inter-animal transforms. *arXiv*.
98. Han, Y. *et al.* (2023) System Identification of Neural Systems: If We Got It Right, Would We Know? in *Proceedings of the 40th International Conference on Machine Learning*, pp. 12430–12444
99. Nonaka, S. *et al.* (2021) Brain hierarchy score: Which deep neural networks are hierarchically brain-like? *iScience* 24, 103013
100. Gröger, F. *et al.* (2026) Revisiting the Platonic Representation Hypothesis: An Aristotelian View. *arXiv*.